JSCIRES

PERSPECTIVE PAPER# Genesis of altmetrics or article-level metrics for measuring efficacy of scholarly communications: Current perspectives

Anup Kumar Das*, Sanjaya Mishra[1]

*Centre for Studies in Science Policy, School of Social Sciences, Jawaharlal Nehru University, [1]Commonwealth Educational Media Centre for Asia, New Delhi, India***ABSTRACT**

The article-level metrics (ALMs) or altmetrics becomes a new trendsetter in recent times for measuring the impact of scientific publications and their social outreach to intended audiences. The popular social networks such as Facebook, Twitter, and Linkedin and social bookmarks such as Mendeley and CiteULike are nowadays widely used for communicating research to larger transnational audiences. In 2012, the San Francisco Declaration on Research Assessment got signed by the scientific and researchers communities across the world. This declaration has given preference to the ALM or altmetrics over traditional but faulty journal impact factor (JIF)-based assessment of career scientists. JIF does not consider impact or influence beyond citations count as this count reflected only through Thomson Reuters' Web of Science® database. Furthermore, JIF provides indicator related to the journal, but not related to a published paper. Thus, altmetrics now becomes an alternative metrics for performance assessment of individual scientists and their contributed scholarly publications. This paper provides a glimpse of genesis of altmetrics in measuring efficacy of scholarly communications and highlights available altmetric tools and social platforms linking altmetric tools, which are widely used in deriving altmetric scores of scholarly publications. The paper thus argues for institutions and policy makers to pay more attention to altmetrics based indicators for evaluation purpose but cautions that proper safeguards and validations are needed before their adoption.

**Keywords:** Altmetrics, article-level metrics, citation database, research assessment, research communication, science communication## INTRODUCTION

In 2014, the Science Citation Index (SCI) – a pioneering product of erstwhile Institute of Scientific Information – completed a journey of 50 years. SCI is considered as a key enabler in making of topical areas of bibliometrics and scientometric. While SCI is completing its 50[th] anniversary, another related area – altmetrics or article-level metrics (ALMs)

*Address for correspondence:
E-mail: anupdas2072@gmail.com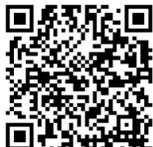

| Access this article online | |
|---|---|
| **Quick Response Code:** | **Website:** www.jscires.org |
| | **DOI:** 10.4103/2320-0057.145622 |is gaining substantial popularity amongst scientific communities, research communicators and research funders. Open access (OA) movement at the beginning of the 21[st] century has strengthened online availability of scholarly publications across the world. The researchers and research communicators attribute the BBB OA declarations as game changers and greater enablers for promotion of scholarly research to larger communities beyond the researchers' fraternities, but also to common citizens and taxpayers. BBB declarations are a group of OA-related declarations namely Budapest declaration in 2002, Berlin declaration in 2003 and Bethesda declaration in 2003 prompting public funded research to be made available and accessible in the public domain. These declarations ensured majority of research publications will be distributed or disseminated through OA channels such as OA knowledge repositories (i.e., green OA channel) and OA journals (i.e., gold OA channel). In this process, a silent transformation also takes place. There is a

82    J Scientometric Res. | May–Aug 2014 | Vol 3 | Issue 2

shift in measuring author's productivity from journal-level indicators to ALM. The citation metrics using journal impact factor (JIF) and Journal immediacy index – both are associated with erstwhile SCI and now Web of Science® and Journal Citation Reports® of Thomson Reuter are felt inadequate in present circumstances while there is increased availability of scholarly publications in online public domain. ALM not only counts citations an individual research papers obtained, but also other influences such as number of downloads, social media share, coverage in news media, etc. The performance measurement for assessing research productivity of individual scientists, as obtained solely from counting number of citations or aggregate/average values of JIF, is no longer valued by funding agencies in developed countries. Rather they started impact evaluation of research publications or funded research projects very differently. Thus, altmetrics of a published paper is measured multi-dimensionally integrating its usage (downloads, views), peer-review (expert opinion), citations, and online interactions (storage, links, bookmarks, conversations).

The San Francisco Declaration on Research Assessment (DORA), singed in 2012 by the scientific and researchers communities across the world, has given preference to the ALM or altmetrics over traditional but faulty JIF-based assessment of career scientists DORA, 2012.[1] The concept of altmetrics explores the potentialities of social media and academic social networks, which helps in increasing global visibility, accessibility and readability of publications shared by the contributing authors Liu, *et al.*, 2014.[2] The researchers in the twenty-first century are very keen to maintain online researchers' profiles in academic social networking websites. They are also interested in transnational networking through online discussion forums and peer-to-peer collaborative platforms. While a plenty of general purpose social networking sites are globally available, some online social networks are meant for academics and researchers. Academic social networking websites facilitate creation of online groups for discussions based on particular research interests. Table 1 in the later part of this paper provides an indicative list of social networking websites that facilitate networking of academics and researchers. All these social networking websites facilitate researchers in building their public profiles – listing their research publications, research projects, research positions or training. While ResearchGate.net, Academia.edu, Linkedin.com, and few others facilitate user-to-user interactions and e-collaborations through e-groups; getCITED.org, SSRN.com and few others do not have such web 2.0 features. Further details of some of these academic social networks are available in the following sub-sections.

## GENESIS AND INSTITUTIONAL FRAMEWORKS

The altmetrics manifesto was published in 2010 by a group of enthusiasts and subsequently it becomes a baseline for a burgeoning altmetrics movement that achieves a global appreciation (Altmetrics.org/manifesto/). In 2011, a dynamic organization was born to technologically support multi-dimensional measurements of published works, beyond citation counts. The name of this start-up company is altmetric LLP, a new avatar in providing online services for generating ALM as a new performance indicator. Simultaneously, the concept of altmetrics is increasingly getting popular since the San Francisco DORA was made public in 2012. Altmetric.com narrates its genesis, as describe below:

"Altmetric LLP was founded by Euan Adie in 2011 and grew out of the burgeoning altmetrics movement.

### Table 1: Major academic social networks

|  | ResearchGate.net | Academia.edu | getCITED.org | SSRN.com |
|---|---|---|---|---|
| Target group | Researchers | Academics: Researchers, students | Researchers | Researchers, authors |
| Founded in | 2008 | 2008 | 2004 | 1994 |
| Subject coverage | All | All | All | Social sciences, humanities and law |
| Mission | To give science back to the people who make it happen and to help researchers build reputation and accelerate scientific progress | To accelerate the world's research; to make science faster and more open | To make records of scholarly work publicly available | To provide rapid worldwide distribution of research to authors and their readers and to facilitate communication among them at the lowest possible cost |
| Account creation | Free | Free | Free | Free |
| Public profile of researchers | Yes | Yes | Yes | Yes |
| Web 2.0 interactivity | Yes | Yes | No | No |

SSRN = Social Science Research Network; getCITED.org = same as getcited.org

Euan had previously worked on Postgenomic.com, an open source scientific blog aggregator founded in 2006. Interested in taking the ideas from postgenomic forward, we entered an altmetrics app into Elsevier's Apps for Science competition and ended up winning. The prize money helped us to grow from an evenings and weekends project into a full-fledged product: The first standalone version of the Altmetric Explorer was released in February 2012. In July 2012, we took on additional investment from Digital Science. Our users now include some of the world's leading journals, funders and institutions. We remain a relatively small company and take pride in our strong focus on engineering and domain knowledge (Source: www.altmetric.com/about.php)."

Since 2013, several scholarly journals and newsletters published special issues on altmetrics. In 2013 the Bulletin of the Association for Information Science and Technology (Bulletin of the ASISandT) published a special issue "altmetrics: What, Why and Where?" with eight articles detailing altmetrics frameworks and possibilities Piwowar, 2013.[3] In the same year, Information Standards Quarterly (ISQ) published a special issue on altmetrics with five articles and an editorial. In June 2014, Research Trends published a special issue "alternative metrics" with nine articles and an editorial (www.researchtrends.com/issue-37-june-2014/) Taylor, *et al.*, 2014.[4] Recently National Information Standards Organization (NISO) of the United States has initiated publishing a whitepaper as an outcome of its ongoing project NISO altmetrics standards project. A draft version of NISO altmetrics Standards Project White Paper got published in May 2014 (See www.niso.org/topics/tl/altmetrics_initiative/). In the same year, Leiden University of the Netherlands publishes a working paper titled Do "altmetrics" Correlate with Citations? Extensive Comparison of altmetric Indicators with Citations from a Multidisciplinary Perspective Fenner, 2013[5] (http://arxiv.org/abs/1401.4321) Costas, *et al.*, 2014.[6]

The CWTS (Center for Science and Technology Studies) of Leiden University has already initiated a research line in altmetrics (www.cwts.nl/Altmetrics), where altmetrics is studied under the umbrella of "Societal Impact of Research". Similarly, LSE (London School of Economics and Political Science, United Kingdom) initiated the LSE Impact of Social Sciences blog in 2011, where different dimensions of ALM are greatly discussed on regular basis (see http://blogs.lse.ac.uk/impactofsocialsciences/tag/altmetrics/). The number of institutions engaged in altmetrics research is growing in western countries.

However, in the global south still there is no formal scholarly research project or research group engaged in-depth studies in this nascent area.

The scholarly publishers, particularly OA publishers, have ridden over altmetrics movement to reach a new height. The Public Library of Science (PLOS) is the most pioneer and early implementer of altmetrics in their OA journals. Every article published in PLOS journals gives instant access to ALM derived from their own algorithms and chosen data sources.

## GROWTH OF LITERATURE ON ALTMETRICS AND ARTICLE-LEVEL METRICS

For the purpose of this paper, the authors have performed an online search in Scopus database using search terms TITLE-ABS-KEY (altmetric*) OR TITLE-ABS-KEY (ALM*), searchable in the title, abstract and keyword fields. The search query retrieved 70 documents as available within Scopus database on July 22, 2014. Retrieved documents were further analyzed to derive year-wise and country-wise distribution of published papers, top contributing authors, and top contributing institutions.

Figure 1 shows the year-wise distribution of papers since the origination or conceptualization of term ALM in 2009. Year 2013 has been most productive in terms of producing literature on altmetrics. Till July 2014, only 14 documents published within year 2014 added to Scopus database. It is expected that more documents will be added for remaining part of year 2014 and will outnumber year 2013.

Figure 2 shows country-wise distribution of papers on the topic of altmetrics. The United States of America stands highest producing 21 papers, United Kingdom stands second with 17 papers and Canada stands third with

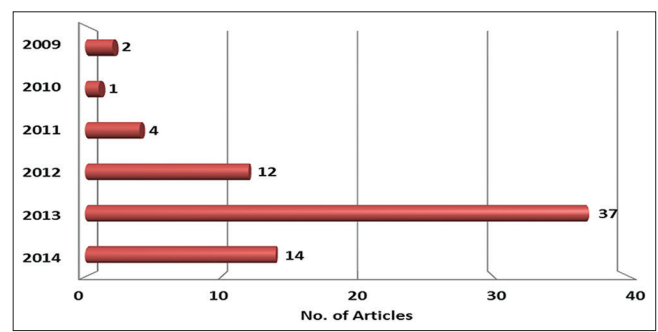

**Figure 1:** Year-wise distribution of Altmetrics Papers (as in Scopus till July 22, 2014)

6 papers. Other contributing countries include Germany, Netherlands, China, Israel, Spain, and Sweden. Countries have one paper each are namely Australia, Austria, Belgium, Brazil, Croatia, Hong Kong, India, Japan, Malaysia, Russian Federation and Switzerland.

Table 2 shows top cited papers on the topic of altmetrics. This table also gives comparative scores of each paper's citations (cited by) statistics as recorded in both in Scopus database as well as Google Scholar (GS) search engine. Paper titled "can tweets predict citations? Metrics of social impact based on Twitter and correlation with traditional metrics of scientific impact", published in 2011, attracted highest number of citations, that is, cited by 59 papers. This paper also attracted 152 citations as recorded in GS database. Paper titled "ALM and the evolution of scientific impact", published in 2009 in PLOS Biology, received the second highest number of citations, that is, cited by 34 papers as in Scopus and 82 papers as in GS.

Table 3 records the top cited papers as retrieved with GS search engine. Publication titled altmetrics: A Manifesto gets the maximum number of citations followed by some papers not covered in Scopus database Priem, *et al.*, 2010.[7] Papers appeared in altmetrics special issues of the Bulletin of the American Society for Information Science and Technology (2013) and ISQ (2013), respectively, started receiving a good number of citations. Interestingly, many of the papers appeared in Tables 2 and 3 are OA contents or freely available online, as indicated in these two tables.

Table 4 shows top contributing authors and their respective affiliation and country. M. Thelwall of United Kingdom contributed the highest number of papers with six contributions on altmetrics topic as recorded in Scopus database, followed by J. Priem of the United States with five contributions. Other authors contributed three papers each. Interestingly, many of them associated with global altmetrics movement and projects related to altmetrics. Table 5 shows top contributing institutions in altmetrics area. The Statistical Cybermetrics Research Group of the University of Wolverhampton, UK is top contributing institution and followed by the School of Information and

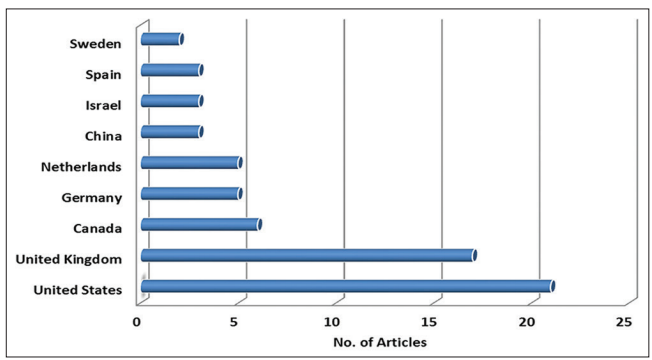

**Figure 2:** Country-wise distribution of Altmetrics Papers (as in Scopus till July 22, 2014)

**Table 2: Top 10 highly cited papers (as in Scopus till 22nd July 2014)**

| Paper details | Cited by | Google scholar citations | Open access article |
|---|---|---|---|
| Eysenbach G. (2011). Can Tweets Predict Citations? Metrics of Social Impact Based on Twitter and Correlation with Traditional Metrics of Scientific Impact. Journal of Medical Internet Research, 13(4) | 59 | 152 | Yes |
| Neylon C. and Wu S. (2009). Article-Level Metrics and the Evolution of Scientific Impact. PLOS Biology, 7(11), e1000242 | 34 | 82 | Yes |
| Piwowar H. (2013). Altmetrics: Value All Research Products. Nature, 493(7431), 159-159 | 16 | 58 | No |
| Thelwall M., Haustein S., Larivière V. and Sugimoto C. R. (2013). Do Altmetrics Work? Twitter and Ten Other Social Web Services. PLOS One, 8(5), e64841 | 14 | 47 | Yes |
| Ware M. (2011). Peer Review: Recent Experience and Future Directions. New Review of Information Networking, 16(1), 23-53 | 14 | 19 | No |
| Priem J., Groth P. and Taraborelli D (2012). The Altmetrics Collection. PLOS One, 7(11), e48753 | 13 | 28 | Yes |
| Yan K. K. and Gerstein M. (2011). The Spread of Scientific Information: Insights from the Web Usage Statistics in PLOS Article-Level Metrics. PLOS One, 6(5), e19917 | 10 | 23 | Yes |
| Schloegl C. and Gorraiz J. (2011). Global usage versus global citation metrics: The case of pharmacology journals. Journal of the American Society for Information Science and Technology, 62(1), 161-170 | 10 | 16 | No |
| Galligan F. and Dyas-Correia S. (2013). Altmetrics: Rethinking the Way We Measure. Serials Review, 39(1), 56-61 | 6 | 18 | No |
| Jacsó P. (2010). Eigenfactor and Article Influence Scores in the Journal Citation Reports. Online Information Review, 34(2), 339-348 | 6 | 12 | No |

PLOS=Public library of science

**Table 3: Other important publications on altmetrics covered in GS search engine**

| Paper details | Google scholar citations | Open access article |
|---|---|---|
| Priem J., Taraborelli D., Groth P. and Neylon, C. (2010). Altmetrics: A Manifesto | 114 | Yes |
| Priem J., Piwowar H. A. and Hemminger B. M. (2012). Altmetrics in the Wild: Using Social Media to Explore Scholarly Impact. arXiv preprint arXiv: 1203.4745 | 67 | Yes |
| Roemer R. C. and Borchardt R. (2012). From Bibliometrics to Altmetrics A Changing Scholarly Landscape. College and Research Libraries News, 73(10), 596-600 | 21 | Yes |
| Adie E. Roe W. (2013). Altmetric: Enriching Scholarly Content with Article-Level Discussion and Metrics. Learned Publishing, 26(1), 11−17 | 18 | Yes |
| Konkiel S. and Scherer D. (2013). New Opportunities for Repositories in the Age of Altmetrics. Bulletin of the ASISandT, 39(4), 22-26 | 15 | Yes |
| Piwowar H. and Priem J. (2013). The Power of Altmetrics on a CV. Bulletin of the ASISandT, 39(4), 10-13 | 15 | Yes |
| Buschman M. and Michalek A. (2013). Are Alternative Metrics Still Alternative?. Bulletin of the ASISandT, 39(4), 35−39 | 13 | Yes |
| Mohammadi E. and Thelwall M. (2014). Mendeley Readership Altmetrics for the Social Sciences and Humanities: Research Evaluation and Knowledge Flows. Journal of the ASISandT, 65(8), 1627-1638 | 13 | No |
| Binfield P. (2009). PLOS One: Background, Future Development, and Article-Level Metrics. Rethinking Electronic Publishing, ELPUB, 69-86 | 12 | Yes |
| Mounce R. (2013). Open Access and Altmetrics: Distinct but Complementary. Bulletin of the ASISandT, 39(4), 14-17 | 11 | Yes |
| Tananbaum G. (2013). Article-Level Metrics: A SPARC Primer. Available from: http://www.sparc.arl.org/sites/default/files/sparc-alm-primer.pdf | 7 | Yes |

GS=Google scholar, PLOS=Public library of science

**Table 4: Top authors in altmetrics (as in Scopus till 22nd July 2014)**

| Name of author | Affiliation | Country | Number of papers |
|---|---|---|---|
| M. Thelwall | Statistical Cybermetrics Research Group, Faculty of Science and Engineering, University of Wolverhampton, Wolverhampton | U.K. | 6 |
| J. Priem | School of Information and Library Science, University of North Carolina at Chapel Hill | USA | 5 |
| P. Groth | Department of Computer Science and The Network Institute, VU University Amsterdam | Netherlands | 3 |
| J. Bar-Ilan | Department of Information Science, Bar-Ilan University, Ramat Gan | Israel | 3 |
| S. Haustein | École de bibliothéconomie et des sciences de l'information, Université de Montréal, Montréal, Canada; b Science-Metrix, Montréal | Canada | 3 |
| I. Peters | Department of Information Science, Heinrich-Heine-University, Düsseldorf | Germany | 3 |
| H. Piwowar | Department of Biology, Duke University | USA | 3 |
| J. Terliesner | Department of Information Science, Heinrich Heine University, Duesseldorf | Germany | 3 |

Library Science of University of North Carolina at Chapel Hill, USA. These institutions are also associated with global altmetrics movement and projects related to altmetrics.

## ALTMETRICS TOOLS

The altmetric LLP remains a pioneer in providing altmetric-related solutions to specifically academic publishers, who would embed altmetric score in each scholarly article they publish in their e-journal gateways. Thus, altmetric score of an online scholarly article is instantly known to visitors of that particular e-journal. In some cases, readers even have convenient options to share bibliographic details of "liked" papers through their social media account. Here, users can instantly share any of these papers through Facebook, Twitter, Google+, Linkedin, Mendeley, CiteULike, or similar interactive social networks.

Figure 3 shows an indicative list of altmetrics tools available to the publishers, funders and researchers. In this figure, symbol "#" denotes that these web services are not comprehensive ones, only provide some aspects of altmetrics. Some web services which have discontinued their experimental beta version of potential altmetric application (but referred in other publications) are not included in this figure, namely ReaderMeter.org, CrowdoMeter.org and ScienceCard.org. Presently, serious contenders of altmetric tools which have much comprehensive approaches are namely Altmetric.com, ImpactStory.org, PlumAnalytics.com and PLOS ALMs. Whereas providers such as PeerEvaluation.org yet to generate a critical mass to be considered as serious contenders of altmetric tools. Some of these tools are also mentioned in the Altmetrics.org/tools/website. Table 6 provides a comparative analysis of major altmetrics providers, namely, Altmetric.com, ImpactStory.org, and PlumAnalytics.com. As indicated in this table, some of the functionalities are common in every platform. These websites provides application programming interface (API) and bookmarklet to publishers and users to fetch altmetric data from different sources. For example, altmetric API of Altmetric.com is an API that enables the publisher to enrich their article pages with ALM data. It helps system to system interaction and obtaining ALM data from different data sources as indicated later. Similarly, altmetric bookmarklet of Altmetric.com is a simple browser tool that lets a researcher instantly gets ALM data for any recent paper. It is a kind of browser plugin that can be integrated into researcher's web browser Chrome, Firefox or Safari.

**Table 5: Top institutions in altmetrics (as in Scopus till 22nd July 2014)**

| Name of institution and country | Number of papers |
| --- | --- |
| Statistical Cybermetrics Research Group, University of Wolverhampton, Wolverhampton, UK | 6 |
| School of Information and Library Science, University of North Carolina at Chapel Hill, USA | 6 |
| Department of Information Science, Bar-Ilan University, Ramat Gan, Israel | 3 |
| VU University, Amsterdam, Netherlands | 3 |
| École de bibliothéconomie et des sciences de l'information, Université de Montréal, Montréal, Canada | 3 |
| Department of Information Science, Heinrich-Heine-University, Düsseldorf, Germany | 3 |
| School of Information and Library Science, Indiana University, USA | 3 |

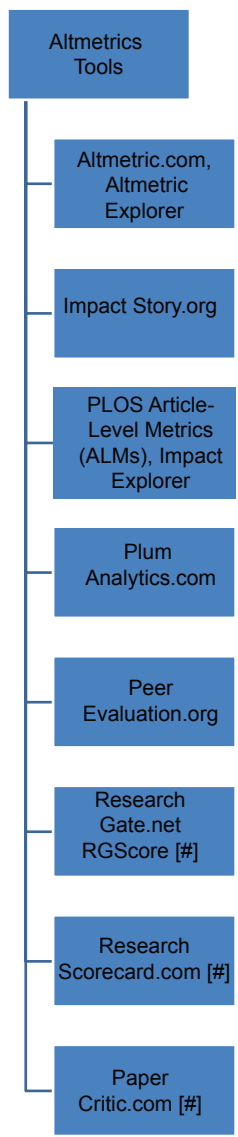

**Figure 3:** Altmetrics tools available to the publishers, funders, and researchers

## DERIVING ALTMETRIC SCORES

### Using Altmetric.com

As indicated in Figure 4, Altmetric Explorer of the Altmetric.com derives altmetric scores from a weighted algorithm covering article-level statistics of viewed, downloaded, cited, saved and discussed. A scholarly article's popularity, usage, acceptance and availability are reflected in an altmetric score. Only articles with digital object identifier (DOI) are considered in arriving at a conclusive altmetric score. Thus, the primary requirement for having an altmetric score is to establish DOI of every published article in electronic journals. Altmetric.com covers about 900+ news sources across the world. Most of them belong to developed or western countries. Few of them belong to developing countries. About 20 news sources are covered from India, namely. The Hindu, Hindustan Times, Times of India, Deccan Herald, Indian Express, the Telegraph, DNA, Asian Age, Business Standard, Dainik Jagran, Dainik Bhaskar, etc., Hence, if a scholarly article is mentioned in any of the news item in 900 + news sources, an artmetric score gets a higher value.

### *Altmetric badge*

Altmetric.com provides a ready-to-use embeddable badge to journal publishers. This badge is embeddable in an article page to help the publishers showcasing impact in a beautiful way. This tool generates small donut shaped multicolor, multilayer visualizations to quickly convey information about each article, with a summary of the score from different data sources. Figure 5 shows an altmetric badge depicting how an article is being outreached and appraised through social media. However, this altmetric score does not include download statistics of the said article.

**Table 6: Major altmetrics providers**

| | **Altmetric.com** | **Impactstory.org** | **Plumanalytics.com** |
|---|---|---|---|
| Target group | Researchers, publishers, librarians, editors, funders | Researchers, publishers, funders | Researchers, publishers, funders |
| Founded in | 2011 | 2012 | 2011 |
| Mission | To track and analyze the online activity around scholarly literature | Discover the full impact of your research | To figure out more accurate ways of assessing research by analyzing the five categories of metrics<br>   Usage<br>   Captures<br>   Mentions<br>   Social media<br>   Citations |
| Functionalities | Authors should be able to see the attention that their articles are receiving in real-time<br><br>Publishers, librarians and repository managers should be able to show authors and readers the conversations surrounding their content<br><br>Editors should be able to quickly identify commentary where a response is required<br><br>Researchers should be able to see which recent papers their peers think are interesting | Researchers who want to know how many times their work has been downloaded, bookmarked, and blogged<br><br>Research groups who want to look at the broad impact of their work and see what has demonstrated interest<br><br>Funders who want to see what sort of impact they may be missing when only considering citations to papers<br><br>Repositories who want to report on how their research products are being discussed<br><br>All of us who believe that people should be rewarded when their work (no matter what the format) makes a positive impact (no matter what the venue)<br><br>Aggregating evidence of impact will facilitate appropriate rewards, thereby encouraging additional openness of useful forms of research output | Assess your impact<br>Track immediate impact<br>Gain an advantage<br>Measure all of your output<br>Group metrics<br>Answer important questions |

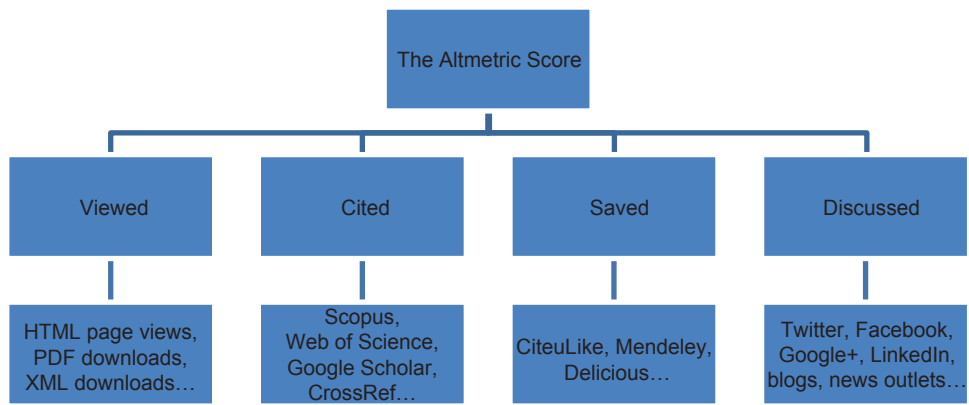

**Figure 4:** Deriving an altmetric score

### Using Public Library of Science Article-level Metric

The PLOS is one of the pioneering publisher that introduced ALM for its OA journals much earlier than many other e-journal gateways. PLOS ALM derived from different data sources as indicated in Figure 6. It includes counts with respect to usage, views, downloads, citations, social bookmarking, blogs, media coverage, and comments. Figure also indicates that PLOS ALM gets view or download statistics not only from PLOS journals but also from the PubMed Central database. Text Box 1 indicates purposes, usages and target users of PLOS ALM. It also helps you understand how PLOS ALM functions. PLOS metrics can be customized to address the needs of researchers, publishers, institutional decision makers, and funders.

### Using ImpactStory.org

The ImpactStory.org is another leading provider of ALM data. This website offers registered users creating their impact profile on the web, revealing diverse impacts of their articles, datasets, software, and more. This is a collaborative not-for-profit open source project supported by the US National Science Foundation, Alfred P. Sloan Foundation and Open Society Foundation. ImpactStory.org helps in creating author's profile and adding publication list through importing bibliographic records from different sources such as Scopus database, ORCID.org, GS citations, SlideShare.net and many others.

A researcher can create a profile for free in this website to know how many times his/her work has been downloaded, bookmarked, and blogged. A researcher can also generate code to embed ImpactStory profile into his institutional CV and the research blog. The homepage of ImpactStory profile of the registered researcher shows a list of contributed papers or presentations. These are categorized as < highly saved>, <highly discussed>, <highly cited>, <saved>, <discussed>, <cited>, and < viewed>. When you click on the title of the paper, you will get a detail ALM score indicating counts from different data sources.

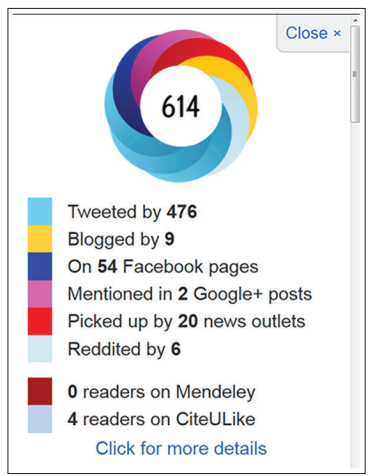

**Figure 5:** An altmetric badge

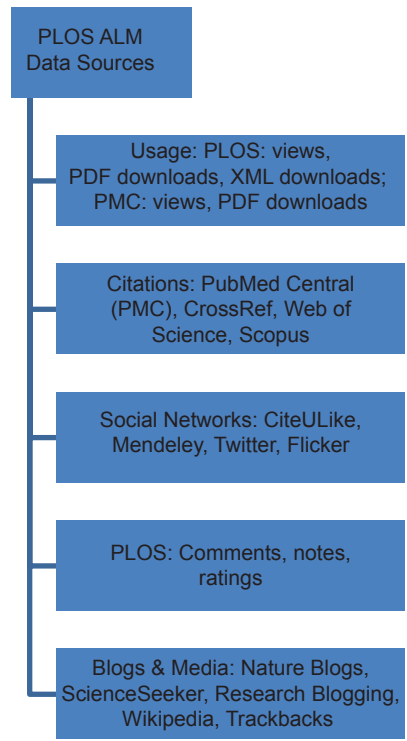

**Figure 6:** Data sources for public library of science article-level metrics

**Text Box 1: Understanding PLOS ALMs**

**PLOS ALMs**

Purpose: ALMs provide a suite of established metrics that measure the overall performance and reach of published research articles

For whom

 Researchers: Maximize the impact of your research

 Publishers: Enhance publication value through real-time views of reach and influence

 Institutions: Capture researcher impact for hiring, tenure, and promotion decisions

 Funders: Track the performance and impact of research funding

ALMs measure the dissemination and reach of published research articles. Traditionally, the impact of research articles has been measured by the publication journal. But a more informative view is one that examines the overall performance and reach of the articles themselves. ALM are a comprehensive set of impact indicators that enable numerous ways to assess and navigate research most relevant to the field itself, including

 Usage

 Citations

 Social bookmarking and dissemination activity

 Media and blog coverage

 Discussion activity and ratings

ALMs are available, upon publication, for every article published by PLOS. Researchers can stay up-to-date with their published work and share information about the impact of their publications with collaborators, funders, institutions, and the research community at large. These metrics are also a powerful way to navigate and discover others' work. Metrics can be customized to address the needs of researchers, publishers, institutional decision-makers, or funders

Source: http://article-level-metrics.plos.org/alm-info/

ALMs=Article-level metrics, PLOS=Public library of science

### Using PlumAnalytics.com

The fourth major altmetric provider is the PlumAnalytics.com. It categorizes metrics into five separate types: Usage, captures, mentions, social media, and citations. PlumAnalytics tracks more than 20 different types of artifacts, including journal articles, books, videos, presentations, conference proceedings, datasets, source code, cases, and more. Figure 7 indicates different data sources used in PlumAnalytics for deriving altmetrics of scholarly publications. As indicated here, PlumAnalytics also includes citation statistics from global patent databases. This is very unique, as compared to other three altmetrics providers. In January 2014, EBSCO has acquired this start-up company Enis, 2014.[8]

### SOCIAL NETWORKS HELPING IMPROVEMENT OF RESEARCHERS' ALTMETRIC SCORES

As we saw in the earlier sections, altmetrics data are derived from various social media and social bookmarking platforms. Researchers of the 21st century need to collaborate with transnational researchers for a successful academic career. They have increased their visibility and participation at the global level through maintaining online profiles, both in general and academic social networking platforms. Their participation in transnational e-groups in online forums, including E-mail-based forums, increased possibilities of peer-to-peer collaborations. While a plenty of general purpose social networking sites are globally available, some online social networks are meant for academics and researchers. Academic social networks facilitate creation of online groups for discussion based on particular research interests. Table 7 provides an indicative list of general purpose social networking websites that also facilitate networking of academics and researchers, besides other citizens. Table 1 provides an indicative list of special purpose websites that mainly facilitate social networking of academics and researchers. While ResearchGate and Academia.edu facilitate user-to-user interactions through e-groups, getCITED.org, and SSRN.com do not have such web 2.0 feature. These academic research networks also ensure peer-to-peer communications through special interest e-groups, where sometimes membership is offered based on prior publications or prior contributions in the research fields

### USING ONLINE SOCIAL BOOKMARKS AND REFERENCE MANAGERS FOR IMPROVING ALTMETRIC SCORES

As we saw in the earlier sections, altmetrics data are also derived from online social bookmarks, and citation or reference managers. Some of the online reference managers also act as PDF organizers, and let others know which papers you read, reviewed or referred to your colleagues. Some of these citation managers also help you to produce subject bibliographies, based on recommended reading lists of your colleagues and e-group members. Thus, online reference managers and social bookmarks play important roles in deciding popularity metrics of research publications accessible online. Individual scholars also get tremendous encouragement when they see their publications are stored, reviewed, recommended and shared by e-groups. There also researchers can create online public profile for highlighting their research publications and reading lists. Table 8 briefly describes major online reference managers and social bookmarks, namely, CiteULike.org, Mendeley.com, Delicious.com and Zotero.org. Zotero is not presently linked to Altmetric.com. Similar few more online reference managers exist, but these are not linked to any altmetric tool used for deriving altmetric score. Some online reference managers, not mentioned in Table 8 although exist, are namely Flow (Flow.proquest.com), EndNote Basic (Endnote.com/basic/), and GS library (Scholar.google.com). Here also users can create an online account for storing references and preparing bibliographies. These are also extensively used by researchers across disciplines. Some reference managers have desktop versions, which are freely downloadable and can be integrated with online accounts. Examples of desktop versions of reference managers are Mendeley and Zotero. EndNote also has a desktop version of the reference manager, although that is not freely available. ProQuest's RefWorks is a reference manager having both online and desktop version. RefWorks' simplified and free version is named Flow, which was launched in 2013 by ProQuest Inc. to be an earnest competitor of the Mendeley, EndNote Basic and Zotero. They will compete each other to increase their market share in the growing segment of online reference managers. Some of them will also be measured for deriving altmetric score of stored or shared research publications.

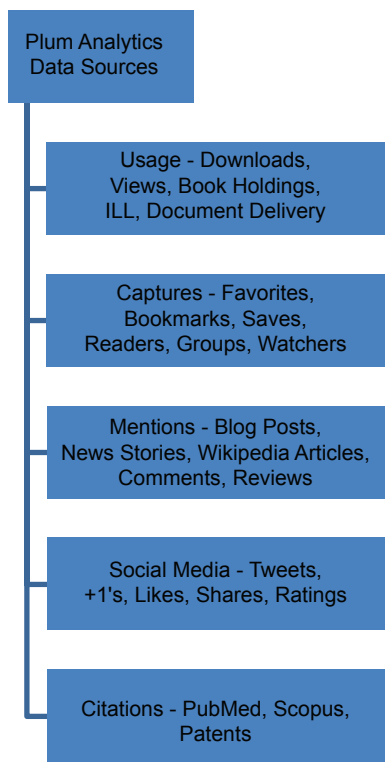

**Figure 7:** Data sources for PlumAnalytics.com article-level metrics

**Table 7: Important general purpose social networks useful for authors and researchers**

|  | Facebook | Twitter | Google+ | LinkedIn | Slideshare | Figshare |
|---|---|---|---|---|---|---|
| Target group | Any citizen | Any citizen | Any citizen | Professionals | Researchers; professionals | Researchers |
| Founded | 2004 | 2006 | 2011 | 2003 | 2006 | 2011 |
| Mission | To give people the power to share and make the world more open and connected | To give everyone the power to create and share ideas and information instantly, without barriers | To bring the nuance and richness of real-life sharing to the web, and making all of Google better by including people, their relationships and their interests | Connect the world's professionals to make them more productive and successful | The world's largest community to share and upload presentations online | Publish all of your research outputs! |
| Public profile of individuals | Yes | Yes | Yes | Yes | Yes | Yes |
| Type of social media | General purpose | General purpose | General purpose | Professional | Format specific | Format specific |
| Acceptable formats | - | - | - | - | Presentations | Datasets, figures and tables |

**Table 8: Major online reference managers and social bookmarks**

|  | CiteULike | Mendeley | Delicious | Zotero |
|---|---|---|---|---|
| Target group | Researchers | Academics: Researchers, students | Researchers, professionals | Researchers |
| Founded in | 2004 | 2008 | 2003 | 2006 |
| About | A free service for managing and discovering scholarly references | A free reference manager and academic social network that can help you organize your research, collaborate with others online, and discover the latest research | Never lose a link again: Delicious is a free and easy tool to save, organize and discover interesting links on the web | A free, easy-to-use tool to help you collect, organize, cite, and share your research sources |
| Ownership |  | Elsevier B.V. | Science Inc. | Center for History and New Media at George Mason University, USA |
| Account creation | Free | Free | Free | Free |

## CONCLUSION

Nowadays the researchers' communities along with research funding agencies are giving much importance to altmetrics, due to better reflection of social impact and outreach of scientific publications using altmetric tools. However, scientific communities in the developing countries are still naïve in handling highly-interactive academic communication channels available to them with web 2.0 readiness. They need to have the necessary information and digital literacy competencies to be conversant with born-digital documents and sharing them with academic social networking platforms. The new-age researchers need to understand and grasp changing landscape of research communications, particularly which are helping global visibility of research communications. To become a successful researcher, one should first become a successful research communicator. One's altmetric score will be increased significantly if he/she manages to reach out to researchers in his/her core and peripheral subject areas using a wide array of social networking platforms available to them. Thus, the nuance of research communication is commensurate with knowledge diffusion to the society.

On the other hand, while discussing efficacy of altmetric indicators and altmetric tools in online discussion groups such as the SIGMETRICS (ASIS and T Special Interest Group on Metrics, sigmetrics@listserv.utk.edu) and the GOAL (Global OA List, goal@eprints.org) during year 2013 to 14, several discussants have pointed out certain limitations or pitfalls of altmetrics. Their concerns are likely: Motivated downloading, automated downloading by special apps or robots, and publisher's inflated downloading data in addition to misuse of traditional citation-based indicators, viz., authors' and journals' self-citation. As social media shares and likes are counted in altmetric scores, researchers and publishers may also push social media shares through undeclared paid advertisements. Although, some discussants do not mind sharing tables of contents in online mailing lists, social media groups, blogs and microblogs, as they considered this as part of research communications strategies adopted by many institutions, individuals, publishers and funders.

Researchers and authors may liberally and ethically use social media tools to boost availability and accessibility of their published works. While other researchers and popular science writers would find the primary research works worthwhile or pioneering, these get mentioned in media articles, science blogs and micro blogs in addition to social bookmarking sites and online reference managers.

Another concern is raised about acquisitions of start-up altmetric providing companies by large corporations. For example, PlumAnalytics.com was acquired by EBSCO LLC in January 2014. This will lead to further commercialization of altmetric business while nonprofit and developing countries' publishers will be marginalized as they have less affordability of portraying altmetric data for every paper they publish. Till today, big publishers and publishers from developed countries are only portraying altmetric data on their respective article page. We need to re-look at these issues before advocating widespread use of altmetric indicators for research assessment.

## ACKNOWLEDGMENTS

This paper is an outcome of a study carried out under the auspices of UNESCO-CEMCA Project titled "Development of curricula and self-directed learning Tools on Open Access." The authors greatly acknowledge supports and professional inputs received from the UNESCO and CEMCA. The authors are also thankful to Dr. Bidyarthi Dutta of DLIS, Vidyasagar University, West Bengal, India for his valuable feedbacks while preparation of this paper.

**How to cite this article:** Das AK, Mishra S. Genesis of altmetrics or article-level metrics for measuring efficacy of scholarly communications: Current perspectives. J Sci Res 2014;3:82-92.

**Source of Support:** Nil, **Conflict of Interest:** None declared